\documentclass[prd,preprintnumbers,
preprint,
superscriptaddress,showpacs,floatfix,nofootinbib,unsortedaddress]{revtex4}
\usepackage{epsfig}
\usepackage{bm}
\usepackage{amssymb}
\usepackage{amsmath}
\usepackage{color}
\usepackage{subfigure}
\newcommand{\be}{\begin{equation}}
\newcommand{\ee}{\end{equation}}
\newcommand{\bea}{\begin{eqnarray}}
\newcommand{\eea}{\end{eqnarray}}

\newcommand{\beq}{\begin{eqnarray}}
\newcommand{\eeq}{\end{eqnarray}}
\newcommand{\nn}{\nonumber \\}

\begin{document}

\title{Analytic solutions of the relativistic Boltzmann equation}

\preprint{YITP-15-11}

\author{Yoshitaka Hatta}
\affiliation{Yukawa Institute for Theoretical Physics, Kyoto University, Kyoto 606-8502, Japan}

\author{Mauricio Martinez}
\affiliation{Department of Physics, The Ohio State University, Columbus, OH 43210, USA}
%\author{Jorge Noronha}
%\affiliation{Instituto de F\'isica, Universidade de S\~ao Paulo, C.P. 66318, 05315-970 S\~ao Paulo, SP, Brazil}
\author{Bo-Wen Xiao}
\affiliation{Key Laboratory of Quark and Lepton Physics (MOE) and Institute
of Particle Physics, Central China Normal University, Wuhan 430079, China}

\date{\today}

\begin{abstract}
We present new analytic solutions to the relativistic Boltzmann equation within the relaxation time approximation. We first obtain spherically expanding solutions which are the kinetic counterparts of the exact solutions of the Israel-Stewart equation in the literature. This allows us to compare the solutions of the kinetic and hydrodynamic equations at an analytical level. We then derive a novel boost-invariant solution of the Boltzmann equation which has an unconventional dependence on the proper time. The existence of such a solution is also suggested in second order hydrodynamics and fluid-gravity correspondence.
\end{abstract}
\maketitle

%%%%%%%%%%%%%%%%%%%%%%%%%%%%%%%%%%%%%%%%%%%%%%%%%%%%%%%%%%%%%%%%%%%%%%%%%%%%%%%%%%%%%%%%%%%
\section{Introduction}
\label{intro}
%%%%%%%%%%%%%%%%%%%%%%%%%%%%%%%%%%%%%%%%%%%%%%%%%%%%%%%%%%%%%%%%%%%%%%%%%%%%%%%%%%%%%%%%%%%

The Boltzmann equation is one of the most important equations in contemporary physics that allows us to characterize the transport properties of a dilute gas based on the microscopic dynamics of the constituent particles \cite{degroot,kremer}. It is a nonlinear partial differential equation for the one-particle distribution function $f(x,p)$ which is very difficult to solve exactly by analytical means. Due to this limitation, it has been more convenient to study the equation by solving it either numerically or by developing approximate solutions based on different expansion schemes. Both approaches  have been successful when comparing their predictions with experimental results, but at the same time, they have their own limitations especially in a relativistic setting. Thus, exact solutions to the Boltzmann equation would certainly be useful to constrain the validity of different perturbative and numerical methods.

Symmetries provide powerful methods to solve and simplify complex problems in physics. This has been particularly useful in relativistic kinetic theory. For instance, for  the relativistic Boltzmann equation within the relaxation time approximation (RTA) \cite{rta}, Refs.~\cite{Denicol:2014xca,Denicol:2014tha} considered a solution having the same symmetry as the Gubser flow \cite{Gubser:2010ze} which is a boost-invariant solution of the conformal  hydrodynamic equations relevant to heavy-ion collisions. The symmetry group of the Gubser flow restricts the number of independent variables as well as their particular combinations
on which the distribution function depends. As a result, the  Boltzmann equation can be effectively reduced to a one-dimensional problem and an exact solution has been constructed  \cite{Denicol:2014xca,Denicol:2014tha}.
%by imposing this specific type of symmetry which renders the mathematical problem effectively one-dimensional~\cite{Denicol:2014xca,Denicol:2014tha}.
By using similar symmetry arguments, other solutions of the RTA Boltzmann equation have been found in the literature for near equilibrium~\cite{Baym:1984np,Gavin:1990up,Heiselberg:1995sh,Wong:1996va,Bialas:1984wv,Bialas:1987en,Banerjee:1989by} and highly anisotropic systems~\cite{Florkowski:2013lza,Florkowski:2013lya,Florkowski:2014sfa,Nopoush:2014qba}. In these results, the solutions are written formally in terms of the effective temperature of the system which has to be determined numerically. While this can be done straightforwardly in practice, it is always welcome to have fully analytical solutions where one can understand various aspects of nonequilibrium dynamics in a completely controllable manner. In this paper, we make progress in this direction by presenting  new analytical solutions to the RTA Boltzmann equation for conformally invariant systems. Each of the solutions is shown to have an explicit counterpart in relativistic viscous hydrodynamics discussed in ~\cite{Hatta:2014gqa,Hatta:2014gga}. This allows us to not only compare the solutions of kinetic and hydrodynamic equations at an analytical level, but also shed light on how the hydrodynamic solutions are obtained as a coarse-grained version of the kinetic solutions.

This paper is organized as follows: In Sect.~\ref{hubblesec} we describe the kinetic theory approach to the Hubble flow solution~\cite{Csorgo:2003rt,Hatta:2014gga}. In Sect.~\ref{sec:Bjorken} we derive a new boost-invariant solution to the RTA Boltzmann equation which features an unusual dependence on the proper time. We provide evidence that such a solution can exist in second order hydrodynamics \cite{Hatta:2014gga} and also in fluid-gravity correspondence \cite{Janik:2010we}. The conclusions of this work are presented in Sect.~\ref{sec:concl}.

%%%%%%%%%%%%%%%%%%%%%%%%%%%%%%%%%%%%%%%%%%%%%%%%%%%%%%%%%%%%%%%%%%%%%%%%%%%%%%%%%%%%%%%%%%%
\section{Kinetic theory description of the Hubble flow}
\label{hubblesec}
%%%%%%%%%%%%%%%%%%%%%%%%%%%%%%%%%%%%%%%%%%%%%%%%%%%%%%%%%%%%%%%%%%%%%%%%%%%%%%%%%%%%%%%%%%%

Our starting point is a spherically expanding solution of the relativistic ideal hydrodynamic equations which is characterized by the following  flow velocity $u^\mu$ and the energy density $\varepsilon$ \cite{Csorgo:2003rt}
 \beq
 u^\mu=\left(\frac{t}{\tau_r},\frac{\vec{r}}{\tau_r}\right)\,, \qquad \varepsilon \propto\frac{1}{\tau_r^4}\,, \label{hubble}
 \eeq
  where $\tau_r \equiv\sqrt{t^2-\vec{r}^2}$ ($\vec{r}=(x_1,x_2,x_3)$) is the  proper time.
 The solution (\ref{hubble}) is valid for the conformal equation of state $\varepsilon=3p$ ($p$ is the pressure) which we assume throughout this paper.
It is convenient to switch from the Minkowski coordinates  to the following coordinate system via a Weyl rescaling
\beq
&&ds^2=-dt^2+dr^2+r^2(d\theta^2+\sin^2\theta d\phi^2) \nn
 \Rightarrow && d\hat{s}^2=\frac{ds^2}{\tau_r^2} = -d\chi^2 + d\upsilon^2 + \sinh^2\upsilon (d\theta^2+\sin^2\theta d\phi^2)\,, \label{coord}
\eeq
 where $\chi = \ln \tau_r$ and we introduced the `rapidity' variable $\upsilon \equiv \tanh^{-1}\frac{r}{t}$. In this coordinate system, the flow velocity is simply $\hat{u}^\mu = \delta^\mu_{\ \chi}$ and the energy density $\hat{\varepsilon}=\tau^4_r\varepsilon$ is a constant. We shall refer to this solution as the  three-dimensional (3D) `Hubble' flow in analogy to the well-known flow solution in cosmology.

Our goal in this section is to describe the 3D Hubble flow and its nonequilibrium generalizations within relativistic kinetic theory. The relativistic Boltzmann equation for the distribution function of massless particles $f(x,p)$ in any curved spacetime reads as~\cite{kremer,degroot}
\beq
%\label{eq:BEcurved}
p^\mu \partial_\mu f + \Gamma_{\mu i}^\lambda p_\lambda p^\mu\, \frac{\partial f}{\partial p_i} = \frac{p^\mu u_\mu}{\tau_{\pi}}\left(f-f_{eq}\right)\,, \label{rta}
\eeq
where we employed the so-called relaxation time approximation (RTA) \cite{rta} in which the collision term is linearized around the equilibrium distribution $f_{eq}$ to be specified shortly. $\tau_{\pi}$ is a characteristic time of the order of the time between succesive collisions and in general, it can depend on the space-time and momentum-space coordinates. In the above equation, $f=f(x^\mu,p_i)$ is considered to be a function of the space-time coordinates $x^\mu$ and the three dimensional spatial momentum components $p_i$ ($i=1,2,3$) with lower (covariant) indices\footnote{The Boltzmann equation~(\ref{rta}) is covariant in coordinate space but not manifestly covariant in momentum-space~\cite{debbasch}. One can sort out this problem by considering a manifestly covariant Boltzmann equation for off-shell distribution functions $f(x^\mu,p_\mu)$~\cite{debbasch2}. We will not consider this approach in our work.}. The energy of the particle is determined from the on-shell condition $g^{\mu\nu}p_\mu p_\nu=0$.

 We analyze the equation (\ref{rta}) in the coordinate system $\hat{x}^\mu=(\chi,\upsilon,\theta,\phi)$. In this case, $\hat{p}_i\equiv (p_\upsilon,p_\theta,p_\phi$), and the on-shell condition becomes
\beq
p_\chi^2 = p_{\upsilon}^2 + \frac{p_{\Omega}^2}{\sinh^2\upsilon}\,, \label{pchi}
\eeq
 where we abbreviated $p_\Omega^2\equiv p_\theta^2+p_\phi^2/\sin^2\theta$.
Computing the Christoffel symbols and using the flow velocity $\hat{u}^\mu=\delta^\mu_\chi$, we find
\beq
\left(p^\chi \partial_\chi + p^\upsilon \partial_\upsilon + p^\theta \partial_\theta+p^\phi\partial_\phi+ \frac{p_{\Omega}^2\cosh \upsilon }{\sinh^3\upsilon} \frac{\partial}{\partial p_\upsilon} + \frac{p_\phi^2\cos\theta }{\sinh^2\upsilon \sin^3\theta}\frac{\partial}{\partial p_\theta} \right)f
= -\beta p^\chi (f-f_{eq})\,.
\label{sym}
\eeq
where we defined $\beta\equiv \tau_r/\tau_\pi$.\footnote{Under the Weyl transformation (\ref{coord}), $\tau_\pi$ is rescaled by a factor of $\tau_r$ \cite{Denicol:2014tha}. }
%It is useful to note that the dimensionless ratio $r_\pi$ differs from $\hat{\tau}_\pi \equiv \tau_\pi \epsilon^{1/4}$ by a factor of constant.
In the following, we assume that $f$ is independent of $\phi$.

First let us specify the equilibrium distribution. Knowing that the flow is static in this coordinate system, we immediately find that the  Boltzmann distribution
\beq
f= f_{eq}(p^\chi) = e^{-p^\chi/\hat{T}}\,,
\eeq
where the temperature $\hat{T}$ is a constant and $p^\chi$ is as in (\ref{pchi}), exactly satisfies (\ref{sym}).\footnote{ We may also take the Bose-Einstein or the Fermi-Dirac distribution as the equilibrium distribution. Actually, in the RTA any function $f_{eq}(p^\chi)$ satisfies (\ref{sym}). We need the full Boltzmann equation to uniquely determine the equilibrium distribution. }
Since $f$ is a scalar invariant, in Minkowski space the equilibrium distribution is
\beq
f_{eq}=e^{-p^\chi/\hat{T}} = e^{-p^\tau/T}\,, \label{hu} %\exp\left(-\frac{1}{T}\sqrt{(p^\upsilon)^2+(p^\theta)^2+(p^\phi)^2}\right)
\eeq
 where $p^\tau=p^\chi/\tau_r$ and $T(\tau_r)=\hat{T}/\tau_r$ is the temperature in Minkowski space. Thus, the distribution function \eqref{hu} is the kinetic counterpart of the ideal Hubble flow.

We now add perturbations on top of the ideal solution. In a conformal theory, $\tau_\pi \propto 1/T$ by dimensional analysis. This means that $\tau_\pi \propto \tau_r$, and therefore $\beta$ in Eq.~\eqref{sym} is a constant. Writing $f=f_{eq}+\delta f= f_{eq}(1+\Phi)$, we find the following equation for $\Phi$
 \beq
\left(p^\chi\partial_\chi+p^\upsilon \partial_\upsilon  +p^\theta\partial_\theta+ \frac{p^2_{\Omega}\cosh\upsilon }{\sinh^3\upsilon}\frac{\partial}{\partial p_\upsilon}+ \frac{p_\phi^2\cos\theta }{\sinh^2\upsilon \sin^3\theta}\frac{\partial}{\partial p_\theta} \right)\Phi
=-\beta p^\chi \Phi\,.
\label{o}
\eeq
We shall solve (\ref{o}) in two interesting cases corresponding to the scalar and tensor perturbations around the ideal solution.

\subsection{Scalar perturbation}

Motivated by a recent work \cite{Noronha:2015jia}, let us  first discuss the solution which is independent of `time' $\chi$. Assuming $O(3)$-symmetry, we find the following exact solution of (\ref{o})
\beq
\Phi&=& {\mathcal K}(p^\chi,p_\Omega)\exp\left(-\frac{1}{2\beta} \ln \frac{p^\chi+p_\upsilon \tanh \upsilon}{p^\chi- p_\upsilon \tanh\upsilon}\right) \nn &=&  {\mathcal K}(p^\chi,p_\Omega)\exp\left(-\frac{1}{\beta} \tanh^{-1}\left(\frac{p_\upsilon \tanh\upsilon}{p^\chi}\right)\right)\,. \label{sim}
\eeq
 The only constraint for the function ${\mathcal K}$ is that it must satisfy the Landau matching condition for the energy density $\varepsilon =\hat{\varepsilon}/\tau_r^4$.
 \beq
 \hat{\varepsilon}= \frac{1}{(2\pi)^3}\int\frac{d^3p}{\sqrt{-g}\,p^\chi}\, (u \cdot p)^2\, f_{eq} =\frac{1}{(2\pi)^3}\int\,\frac{d^3p}{\sqrt{-g}\,p^\chi}\, (u \cdot p)^2\, f\,,
%= \frac{3}{\pi^2}\hat{T}^4\,.
\label{energy0}
\eeq
 where $d^3p=dp_\upsilon dp_\theta dp_\phi$.
  If we further assume that ${\mathcal K}$ does not depend explicitly on $p_\Omega$, then the condition reduces to
 \beq
 \int_0^\infty dp^\chi p_\chi^3 e^{-p_\chi/\hat{T}} {\mathcal K}(p^\chi)=0\,.
 \eeq
   The above solution (\ref{sim}) has been derived in an analogous way to \cite{Noronha:2015jia} based on a different ideal hydrodynamic solution. As discussed in that reference,  this type of solutions is characterized by certain scalar moments of $f$ and the associated entropy production despite the vanishing shear-stress tensor
\beq
\hat{\pi}^{\mu\nu} = \frac{1}{(2\pi)^3}\int\frac{ d^3p}{\sqrt{-g}\,p^\chi}\, \Delta^{\mu\nu}_{\alpha\beta}p^\alpha p^\beta \delta f=0\,, \label{shear}
\eeq
 where $\Delta^{\mu\nu}_{\alpha\beta}=\frac{1}{2}(\Delta^\mu_\alpha \Delta^\nu_\beta + \Delta^\mu_\beta \Delta^\nu_\alpha)-\frac{1}{3}\Delta^{\mu\nu}\Delta_{\alpha\beta}$ with $\Delta^{\mu\nu}=g^{\mu\nu}+u^\mu u^\nu$.
 (\ref{shear}) means that the solution does not allow for a hydrodynamic description. Rather, it represents the relaxation of `fast' degrees of freedom usually not taken into account in hydrodynamics.

\subsection{Tensor perturbation}

We now return to (\ref{o}) and consider the tensor perturbations.
We parameterize the nonequilibrium part in such a way that the connection to hydrodynamics is transparent
\beq
\Phi= \frac{\pi^2}{8\hat{T}^6}  p^\mu p^\nu \hat{\pi}_{\mu\nu}(\chi,\upsilon,\theta)\equiv e^{-\beta\chi }\frac{\pi^2}{8\hat{T}^6}  p^\mu p^\nu \widetilde{\pi}_{\mu\nu}(\upsilon,\theta)\,. \label{part}
\eeq
The first equality is the standard parameterization in the moment method where $\hat{\pi}^{\mu\nu}$ is identified with the  shear stress tensor in viscous hydrodynamics (\ref{shear}). In the second equality, we extracted the exponential relaxation factor $e^{-\beta \chi}$ which turns into a power-law behavior $\tau_r^{-\beta}$ in the original Minkowski space.

 Let us first look for $O(3)$-symmetric solutions where $\Phi$ is independent of $\theta$ and $\hat{\pi}^{\theta}_{\ \theta}=\hat{\pi}^{\phi}_{\ \phi}=-\frac{1}{2}\hat{\pi}^{\upsilon}_{\ \upsilon}$.
 Substituting Eq.~\eqref{part} in Eq.~\eqref{o}, we find the following equation for $\widetilde{\pi}^{\upsilon}_{\ \upsilon}$
\beq
%&&\left( p^\upsilon \partial_y  +p_{\Omega}^2 \frac{\coshy }{\sinh^3y}\frac{\partial}{\partial p_y} \right)p^\mu p^\nu \tilde{\pi}_{\mu\nu}(y) \nn
\left( p^\upsilon \partial_\upsilon  +p_{\Omega}^2 \frac{\cosh\upsilon }{\sinh^3\upsilon}\frac{\partial}{\partial p_\upsilon} \right) \left(p_\upsilon^2-\frac{p_\Omega^2}{2\sinh^2\upsilon}\right)\widetilde{\pi}^\upsilon_{\ \upsilon}(\upsilon) =0\,. \label{uu}
\eeq
 However, we immediately encounter a difficulty. It is easy to see that there can be no solution to (\ref{uu}). Indeed, the general solution of the differential equation
 \beq
 \left( p^\upsilon \partial_\upsilon  +p_{\Omega}^2 \frac{\cosh\upsilon }{\sinh^3\upsilon}\frac{\partial}{\partial p_\upsilon} \right)\Phi=0\,,
 \eeq
 is  $\Phi=\Phi(p_\chi,p_\Omega)$, and this is clearly incompatible with the structure of  (\ref{uu}).

The trouble is that this negative result is in apparent contradiction to the finding in Ref.~\cite{Hatta:2014gga}. There, the authors derived exact solutions to the Israel-Stewart equation in hydrodynamics which relax to the Hubble flow at large times.
%\beq
%&&\quad 4\varepsilon u^\nu \nabla_\nu u^\mu + \Delta^{\mu\alpha}\nabla_\alpha \varepsilon+3\Delta^\mu_{\ \nu}\nabla_\alpha \pi^{\alpha\nu}=0\,, \label{first}\\
%&& \pi^{\mu\nu}=-2\eta \sigma^{\mu\nu}-\tau_\pi \left(\Delta^{\mu}_\alpha \Delta^\nu_\beta u^\rho \nabla_\rho \pi^{\alpha\beta}+\frac{4}{3}\pi^{\mu\nu} \vartheta  \right)\,. \label{pi}
%\eeq
In the present notation, the $O(3)$-symmetric solution is (see (74) of \cite{Hatta:2014gga})
\beq
\widetilde{\pi}^\upsilon_{\ \upsilon} \propto \frac{1}{\sinh^3\upsilon}\,, \label{we}
\eeq
%but we have just seen that this
which, however, is not a solution to (\ref{uu}) as we have just seen in the above derivation. On general grounds, it is expected that every solution of the hydrodynamic equations has a microscopic counterpart in kinetic theory,
and actually our motivation here is to rederive (\ref{we}) as the solution of the Boltzmann equation.
It is tempting to think that this is a problem of the RTA which oversimplifies the collision term of the Boltzmann equation and therefore restricts the solution space of kinetic theory. Yet, one can derive the Israel-Stewart equation starting from the Boltzmann equation in the RTA (see for example~\cite{Jaiswal:2013npa}), and this implies that the above conflict must somehow be reconciled within the RTA.

In order to understand in what sense Eq.~\eqref{we} is a solution, let us substitute it into Eq.~\eqref{uu}
\beq
\left( p_\upsilon \partial_\upsilon  +p_{\Omega}^2 \frac{\cosh\upsilon }{\sinh^3\upsilon}\frac{\partial}{\partial p_\upsilon} \right)\! \left(p_\upsilon^2-\frac{p_\Omega^2}{2\sinh^2\upsilon}\right)\frac{1}{\sinh^A \upsilon} \!=
\frac{-p_\upsilon \cosh\upsilon}{2\sinh^{A+1}\upsilon} \left(2A p_\upsilon^2 - \frac{(A+6)p_\Omega^2}{\sinh^2\upsilon}\right)\!\,,  \label{non}
\eeq
 where $A=3$. As expected, the right-hand-side does not vanish for any value of $A$. However, the question is whether these unwanted terms affect the hydrodynamic equations. The energy-momentum conservation equation is
 \beq
 0=\nabla_\nu T^{\mu\nu} =\frac{1}{(2\pi)^3} \int \frac{d^3p}{\sqrt{-g}p^\chi}p^\mu p^\nu\left(\partial_\nu+ \Gamma_{\nu i}^\lambda p_\lambda\frac{\partial}{\partial p_i}\right)(f_{eq}+\delta f)\,. \label{conservation}
 \eeq
 Taking the component $\mu=\upsilon$, we see that all that is needed for the hydrodynamic equation to hold is that the following integral vanishes
 \beq
 \int \frac{d^3p}{\sqrt{-g}p^\chi}p_\upsilon \left(\mbox{Eq}.(\ref{non})\right)\propto \int \frac{d^3p}{\sqrt{-g}p^\chi}p_\upsilon^2\left(2Ap_\upsilon^2-\frac{(A+6)p_\Omega^2}{\sinh^2y}\right)\propto A-3\,.
 \eeq
 This is indeed the case when $A=3$. It is easy to see that the components of the equation (\ref{conservation}) other than $\mu=\upsilon$ are trivially satisfied even when $A\neq 3$.

The above analysis teaches an important lesson about comparing solutions of kinetic and hydrodynamic equations. Since the hydrodynamic equation is a course-grained version  of the kinetic equation, it admits a class of solutions which are not sensitive to the exact details of kinetic theory. To accommodate this, we have enlarged the solution space of the Boltzmann equation in the RTA to allow the equation to be satisfied up to terms that do not affect the macroscopic (hydrodynamic) equations. With this qualification, remarkably the solutions of the Boltzmann and hydrodynamic equations are exactly the same in the sense that they are characterized by the same macroscopic variables $\varepsilon$, $\pi^{\mu\nu}$, etc. This is in contrast to the common perception that solutions of the hydrodynamic equation are only an approximate version of the solutions of the Boltzmann equation, as repeatedly observed in the literature \cite{Huovinen:2008te,Jaiswal:2013npa,Florkowski:2013lza,Florkowski:2013lya,Denicol:2014xca,Denicol:2014tha}.\\

In Ref.~\cite{Hatta:2014gga}, along with the $O(3)$-symmetric solution~\eqref{we},  non-$O(3)$-symmetric solutions  to the Israel-Stewart equation were also obtained. It is straightforward to generalize the present analysis to this case. Here we consider only one of the non-$O(3)$-invariant solutions found in \cite{Hatta:2014gga} which reads\footnote{We have checked that the other non-$O(3)$-symmetric solution in \cite{Hatta:2014gga}, which is proportional to $\frac{1}{\sin^{3/2}\theta \sinh^3\upsilon}$, also satisfies the RTA Boltzmann equation following exactly the same pattern.}
\beq
\widetilde{\pi}^{\upsilon}_{\ \upsilon}=\widetilde{\pi}^{\theta}_{\ \theta}=-\frac{1}{2}\widetilde{\pi}^{\phi}_{\ \phi}\propto \frac{1}{\sin^3\theta \sinh^3\upsilon}\,, \label{sol}
\eeq
so that
\beq
p^\mu p^\nu \widetilde{\pi}_{\mu\nu}(\upsilon,\theta)\propto \left(p_\upsilon^2+\frac{p_\theta^2}{\sinh^2\upsilon}-\frac{2p_{\phi}^2}{\sinh^2\upsilon \sin^2\theta}\right)\frac{1}{\sinh^3\upsilon \sin^3\theta}\,. \label{ag}
\eeq
Substituting (\ref{ag}) into (\ref{o}), we find
\beq
&& \left(p^\upsilon \partial_\upsilon  +p^\theta\partial_\theta+ \frac{p^2_{\Omega}\cosh\upsilon }{\sinh^3\upsilon}\frac{\partial}{\partial p_\upsilon}+ \frac{p_\phi^2\cos\theta }{\sinh^2\upsilon \sin^3\theta}\frac{\partial}{\partial p_\theta} \right)p^\mu p^\nu \widetilde{\pi}_{\mu\nu}(\upsilon,\theta) \nn
&& \quad \propto \frac{-3}{\sin^3\theta \sinh^4\upsilon}\left(p_\upsilon \cosh\upsilon + \frac{p_\theta \cot\theta}{\sinh \upsilon}\right)
\left(p_\upsilon^2+\frac{p_\theta^2}{\sinh^2\upsilon}-\frac{4p_\phi^2}{\sinh^2\upsilon
\sin^2\theta} \right)\,. \label{term}
\eeq
As it happened in the $O(3)$-symmetric case, the right hand side of (\ref{term}) does not vanish. However, now we know how to sort out this. The nonvanishing terms in Eq.~(\ref{term}) do not affect the hydrodynamic equation (\ref{conservation}) because
   \beq
 \int \frac{d^3p}{\sqrt{-g}p^\chi}p_\upsilon \left(\mbox{Eq}.(\ref{term})\right)= \int \frac{d^3p}{\sqrt{-g}p^\chi}p_\theta \left(\mbox{Eq}.(\ref{term})\right)=0\,,
 \eeq
  for $\mu=\upsilon, \theta$. (The other components are trivial.)
 In this sense, Eq.~\eqref{ag} is the solution to the Boltzmann equation which corresponds to the hydrodynamic solution~\eqref{sol}.

Finally, we note that in both the $O(3)$-invariant and non-invariant cases one can obtain the free streaming solutions by taking the limit $\beta\,\to\,0$
\beq
f=f_{eq}\left(1+\frac{\pi^2}{8\hat{T}^6}  p^\mu p^\nu \widetilde{\pi}_{\mu\nu}(\upsilon,\theta) \right)\,.
\eeq
In the absence of the exponential damping, the system never reaches thermal equilibrium (see, also, \cite{Denicol:2014tha,Calzetta:2014hra}).

%%%%%%%%%%%%%%%%%%%%%%%%%%%%%%%%%%%%%%%%%%%%%%%%%%%%%%%%%%%%%%%%%%%%%%%%%%%%%%%%%%%%%%%%%%%
\section{Bjorken flow revisited}
\label{sec:Bjorken}
%%%%%%%%%%%%%%%%%%%%%%%%%%%%%%%%%%%%%%%%%%%%%%%%%%%%%%%%%%%%%%%%%%%%%%%%%%%%%%%%%%%%%%%%%%%

In the previous section, we presented explicit analytic solutions of the Boltzmann equations in the RTA. This has been possible largely due to the fact that the ideal hydrodynamic solution is  static in a cleverly chosen coordinate system.  For essentially nonstatic flows, analytic solutions of the Boltzmann equation are difficult to obtain even in the RTA. Nevertheless, in the boost-invariant case relevant to heavy-ion collisions \cite{Bjorken:1982qr}, one can gain analytic insights into the behavior of the flow at late times. In this section we revisit this problem and elucidate a new solution of the Boltzmann equation which exists in the presence of conformal symmetry.

We work in the coordinate system
\beq
ds^2=-d\tau^2+\tau^2d\zeta^2+dx_T^2 +x_T^2d\phi^2\,, \label{boostc}
\eeq
 where $x_T=\sqrt{x_1^2+x_2^2}$ is the transverse coordinate. $\tau=\sqrt{t^2-x_3^2}$ and $\zeta=\tanh^{-1}\frac{x_3}{t}$ are the one-dimensional analogs of the three-dimensional proper time $\tau_r$ and the rapidity $\upsilon$ introduced in the previous section.  The RTA Boltzmann equation with the comoving flow velocity $u^\mu=\delta^\mu_\tau$ \cite{Bjorken:1982qr} was first studied in  \cite{Baym:1984np} and developed more recently in \cite{Gavin:1990up,Heiselberg:1995sh,Wong:1996va,Bialas:1984wv,Bialas:1987en,
 Banerjee:1989by,Florkowski:2013lza,Florkowski:2013lya}. Assuming that $f$ depends only on $\tau$, we need to solve
\beq
\partial_\tau f= -\frac{1}{\tau_\pi} (f-f_{eq})\,. \label{bjorken}
\eeq
  The equilibrium distribution is taken to be the Boltzmann distribution as before $f_{eq}=e^{u\cdot p/T}=e^{-p^\tau/T(\tau)}$ where $p^\tau = \sqrt{ p_\zeta^2/\tau^2+p_T^2}\,$, but unlike the Hubble flow case, $T(\tau)$ is an unknown function which is dynamically determined from the first moment of the Boltzmann equation~\eqref{bjorken} which gives us the dynamical Landau matching condition~\cite{Baym:1984np}
   \beq
 \varepsilon= \frac{1}{(2\pi)^3}\int\frac{d^3p}{\sqrt{-g}\,p^\tau}\, (u \cdot p)^2\, f_{eq} =\frac{1}{(2\pi)^3}\int\,\frac{d^3p}{\sqrt{-g}\,p^\tau}\, (u \cdot p)^2\, f\,.
\label{energy}
\eeq
Consequently, $f=f_{eq}$ does not solve (\ref{bjorken}) exactly, which is a manifestation of the nonstatic nature of the geometry.

In terms of the energy density (\ref{energy}), the solution of (\ref{bjorken}) is formally given by \cite{Baym:1984np,Florkowski:2013lya}
\beq
\varepsilon(\tau) = D(\tau,\tau_0)\varepsilon_0(\tau) + \int^\tau_{\tau_0}\frac{d\tau'}{\tau_\pi(\tau')}D(\tau,\tau')\varepsilon(\tau')\frac{1}{2}\left(\frac{\tau'^2}{\tau^2}+\frac{\arctan \sqrt{\frac{\tau^2}{\tau'^2}-1}}{\sqrt{\frac{\tau^2}{\tau'^2}-1}}\right)\,,
\label{yo}
\eeq
where
\beq
D(\tau,\tau_0)=\exp\left(-\int^\tau_{\tau_0}\frac{d\tau'}{\tau_\pi(\tau')}\right)\,,
\eeq
and
\beq
\varepsilon_0(\tau)=\frac{1}{(2\pi)^3}\int \frac{d^3p}{\tau p^\tau} (p^\tau)^2 f(\tau_0,p_\zeta, p_T)\,. \label{e0}
\eeq
In Eq.~\eqref{e0}, $f(\tau_0)$ is the initial distribution at $\tau_0\, $, whereas $p^\tau = \sqrt{p^2_\zeta/\tau^2+p_T^2}$ (on-shell condition) is defined at $\tau$.
After integration by parts, Eq.~\eqref{yo} can be written as
\beq
\int_{\tau_0}^\tau d\tau' D(\tau,\tau')\frac{\partial}{\partial \tau'}\left(\varepsilon(\tau')R\left(\frac{\tau'}{\tau}\right)\right) = D(\tau,\tau_0)\left(\varepsilon_0(\tau)-\varepsilon(\tau_0)R
\left(\frac{\tau_0}{\tau}\right)\right)\,, \label{baym}
\eeq
 where we abbreviated
 \beq
 R(x)\equiv\frac{1}{2}\left(x^2+\frac{\arctan \sqrt{\frac{1}{x^2}-1}}{\sqrt{\frac{1}{x^2}-1}}\right)\,.
 \eeq

 Following \cite{Baym:1984np}, let us first assume that $\tau_\pi$ is a constant. Then $D(\tau,\tau')=e^{-(\tau-\tau')/\tau_\pi}$, and (\ref{baym}) becomes
\beq
\int_{\tau_0}^\tau d\tau' e^{\tau'/\tau_\pi}\frac{\partial}{\partial \tau'}\left(\varepsilon(\tau')R\left(\frac{\tau'}{\tau}\right)\right) = e^{\tau_0/\tau_\pi}\left(\varepsilon_0(\tau)-
\varepsilon(\tau_0)R\left(\frac{\tau_0}{\tau}\right)\right)\,. \label{contrast}
\eeq
Since the right-hand side decreases with time as ${\mathcal O}(1/\tau)$, one should not allow the left-hand side to  grow exponentially in $\tau$. This leads to the condition
\beq
\left.\frac{\partial}{\partial \tau'}\left(\varepsilon(\tau')R\left(\frac{\tau'}{\tau}\right)\right) \right|_{\tau'=\tau}=0\,.
\label{cond}
\eeq
With $R(1)=1$ and $R'(1)=4/3$, (\ref{cond}) gives  $\varepsilon(\tau) \sim 1/\tau^{4/3}$. Thus the Bjorken solution is recovered  \cite{Bjorken:1982qr}. The same conclusion is reached if $\tau_\pi$ depends on time as $\tau_\pi(\tau) \sim \tau^p$ with $0\le p <1$. However, when $p=1$, this argument breaks down.\footnote{The case $p=1$ was previously studied in \cite{Gavin:1990up} without assuming conformal symmetry. It was found that the Bjorken solution $\varepsilon\sim 1/\tau^{4/3}$ is recovered only in the limit $\beta\to \infty$ (see Eq.~\eqref{beta}), whereas for other values of $\beta$ the asymptotic behavior is $\varepsilon \sim 1/\tau^\gamma$ with $1<\gamma<\frac{4}{3}$.  The special solution (\ref{cc}) which we are going to derive was not noticed in \cite{Gavin:1990up}.} In fact, there exists a novel asymptotic solution of the form
\beq
\varepsilon(\tau)\approx  \frac{C}{\tau^4}\,, \quad (\tau\gg \tau_0) \label{cc}
\eeq
 whose normalization constant $C$ is not arbitrary but is an intrinsic parameter of a given theory.

In order to verify this statement, we first note that the two conditions $p=1$ and $\varepsilon\sim 1/\tau^4$ are naturally related in the presence of conformal symmetry.  Indeed, in a conformal theory, $\tau_\pi \propto \varepsilon^{-1/4}$ by dimensional analysis. We can then write in the asymptotic regime the following
\beq
\tau_\pi\equiv \frac{\hat{\tau}_\pi}{\varepsilon^{1/4}} \approx \frac{\hat{\tau}_\pi}{C^{1/4}}\tau \equiv\frac{\tau}{\beta} \,, \label{beta}
\eeq
so that $p=1$ together with
\beq
 D(\tau,\tau')=\left(\frac{\tau'}{\tau}\right)^{\beta}\,.
\eeq
 lead us to write Eq.~(\ref{baym}) as
\beq
\int_{\tau_0}^\tau d\tau' \tau'^{\beta}\frac{\partial}{\partial \tau'}\left(\varepsilon(\tau')R\left(\frac{\tau'}{\tau}\right)\right) = \tau_0^{\beta}\left(\varepsilon_0(\tau)-\varepsilon(\tau_0)R\left(\frac{\tau_0}{\tau}\right)\right)\,. \label{care}
\eeq
We see that, instead of an exponentially growing factor as in (\ref{contrast}), the integrand contains only powers of $\tau'$. At large times $\tau\gg \tau_0$,  the right hand side is of order ${\mathcal O}(1/\tau)+{\mathcal O}(1/\tau^3)$ with $\tau_0$-dependent coefficients. This should match the contribution from the lower bound $\tau'=\tau_0$ of the $\tau'$-integration on the left hand side. Then the contribution from the upper bound $\tau'=\tau$ of order ${\mathcal O}(1/\tau^{4-\beta})$, which is independent of $\tau_0\,$, must vanish.

For generic values of $\beta$, the $\tau'$-integral in Eq.~(\ref{care}) cannot be done analytically.
 In order to study the behavior near the upper limit, we expand $R(\tau'/\tau)$ in powers of $\tau-\tau'$
 \beq
 R\left(\frac{\tau'}{\tau}\right)=1+\frac{4}{3}\left(\frac{\tau'}{\tau}-1\right)+\frac{2}{5}
 \left(\frac{\tau'}{\tau}-1\right)^2+\cdots\,.
 \eeq
 and integrate over $\tau'$ term by term using $\varepsilon(\tau')\sim 1/\tau'^4$.
 In fact, we need to expand to all orders in $\tau-\tau'$.
%\beq
%C\int^\tau d\tau' \tau'^{a} \left(-\frac{4}{\tau'^5} + \frac{4}{3\tau\tau'^4} \right) \approx \frac{4}{3}C \tau^{a -4}\left(\frac{1}{a-3} -\frac{3}{a-4}\right) \label{um}
%\eeq
% Requiring this to vanish, we get $a=\frac{5}{2}$ so that $C^{1/4}=5\hat{\tau}_\pi/2$.
 In practice, we used  Mathematica and  expanded $R(\tau'/\tau)$ to ${\mathcal O}((\tau-\tau')^{50})$. We then require that the contribution from the upper bound $\tau'=\tau$ vanishes. This yields the value\footnote{In addition to this, Mathematica finds other positive roots such as $\beta=2$ and $\beta=4$. We discard them as artifacts. For these values of $\beta$, one can evaluate the $\tau'$-integral explicitly, and find inconsistencies with (\ref{care}).}
 \beq
\beta \approx 1.27672\,. \label{aa}
 \eeq
  In Appendix \ref{app}, we present another derivation of this constant.
 We thus conclude that, in the presence of conformal symmetry, the Boltzmann equation in the RTA admits the following asymptotic solution
\beq
\varepsilon=\frac{C}{\tau^4}\, \qquad \textrm{with} \quad C^{1/4} \approx 1.28\hat{\tau}_\pi\,. \label{rt}
%1.27672\hat{\tau}_\pi\,. \label{rt}
\eeq
As announced, the normalization is completely determined by $\hat{\tau}_\pi=\tau_\pi \varepsilon^{1/4}$ which is an intrinsic parameter of a given theory. Note that in QCD at high temperature which is nearly conformal, we have $\tau_\pi \sim \eta/\varepsilon \sim (T\lambda^2 \ln \frac{1}{\lambda})^{-1}$ ($\lambda=g^2N_c$ is the 't Hooft coupling)  \cite{Arnold:2000dr} so that
\beq
\varepsilon \sim \frac{N_c^2}{\left(\tau \lambda^2 \ln \frac{1}{\lambda}\right)^4}\,.
\eeq

We now discuss the connection to hydrodynamics.
In fact, a  solution very similar to Eq.~\eqref{rt} has been previously found in Ref.~\cite{Hatta:2014gga} as an exact solution of the general second-order hydrodynamic equations and dubbed `unorthodox Bjorken flow'. This is given by (see Eq.~(112) of Ref.~\cite{Hatta:2014gga})
 \beq
 \varepsilon=\frac{C}{\tau^4}\,\qquad   \textrm{with} \quad C^{1/4}=\frac{16\hat{\tau}_\pi-3\hat{\eta}-2\hat{\tau}_{\pi\pi}}{6}\,,
 \label{ortho}
 \eeq
where $\eta=\varepsilon^{3/4}\hat{\eta}$ is the shear viscosity and $\tau_{\pi\pi}=\varepsilon^{-1/4}\hat{\tau}_{\pi\pi}$ is one of the second-order transport coefficients in the constitutive equation for $\pi^{\mu\nu}$
\beq
\pi^{\mu\nu}=-2\eta \sigma^{\mu\nu}-\tau_\pi \Delta^{\mu}_{\alpha}\Delta^\nu_\beta \nabla_\tau \pi^{\alpha\beta}-\tau_{\pi\pi}\Delta^{\mu\nu}_{\alpha\beta}
\sigma^{\alpha\lambda}\pi^{\beta}_{\ \lambda}+\cdots\,.
 \eeq
 These transport coefficients can be evaluated from the Boltzmann equation in the RTA.\footnote{The other second-order transport coefficients which are not derivable from the Boltzmann equation in the RTA have been ignored in (\ref{ortho}).}
  Using $\tau_\pi=5\eta/(Ts)=15\eta/(4\varepsilon)$ from the Chapman-Enskog theory ($s$ is the entropy density) and $\tau_{\pi\pi}=10\,\tau_\pi/7$ for the massless Boltzmann gas  \cite{Jaiswal:2013npa,Denicol:2012cn}, we get
\beq
C^{1/4}=\beta \,\hat{\tau}_\pi\approx 2.06\,\hat{\tau}_\pi\,, \label{can}
%2.05714\hat{\tau}_\pi\,,
\eeq
 which is in the same ballpark as Eq.~\eqref{rt}. The discrepancy may be alleviated in a more precise evaluation of these coefficients from the Boltzmann equation.
Conversely, if we assume that the Chapman-Enskog value for $\eta$ is precise, our result may be used to estimate the value of $\tau_{\pi\pi}$ in the RTA. Equating (\ref{rt}) and (\ref{ortho}), we obtain
%\beq
$\tau_{\pi\pi}\approx 3.77\,\tau_\pi\,$,
%3.76984\tau_\pi\,.
%\eeq
 which turns out to be a few times larger than the results previously derived by different authors~\cite{Denicol:2012cn,Jaiswal:2013npa}.

\subsection{Fluid-Gravity duality}

In this subsection, we point out that the existence of the solution of the type (\ref{rt}) is also suggested in the framework of fluid-gravity correspondence.\footnote{This subsection is largely motivated by interesting discussions with Jorge Noronha to whom we are grateful.} In strongly coupled ${\mathcal N}=4$ supersymmetric Yang-Mills theory, hydrodynamic flows are dual to solutions of the five-dimensional (5D) Einstein equation in asymptotically Anti-de Sitter (AdS) spaces (see Ref.~\cite{Janik:2010we} for a pedagogical review)
\begin{equation}
R_{AB}-\frac{1}{2}G_{AB}R-6G_{AB}=0\,, \label{ein}
\end{equation}
where $R_{AB}$ and $R$ are the Ricci tensor and the scalar curvature, respectively. In the Fefferman-Graham coordinates, the 5D metric $G_{AB}$ can be written as
\beq
ds_5^2 = G_{AB}dx^A dx^B=\frac{1}{z^2}\left[g_{\mu\nu}(x,z)dx^{\mu}dx^{\nu}+dz^2\right]\,,
\eeq
 where $z$ is the `fifth' dimension. Near the Minkowski boundary $z=0$, the solution of (\ref{ein}) takes the form
\begin{equation}
g_{\mu\nu} =g_{\mu\nu}^{(0)} +z^4 g_{\mu\nu}^{(4)}+ z^6 g_{\mu\nu}^{(6)} +\cdots\,,
\end{equation}
where $g_{\mu\nu}^{(0)}$ is the four dimensional flat Minkowski metric and $g_{\mu\nu}^{(4)}$ is related to the energy momentum tensor
of in the gauge theory as
\begin{equation}
T_{\mu\nu} =\frac{N_c^2}{2\pi^2} g_{\mu\nu}^{(4)}\,.
\end{equation}

 For boost-invariant flows, the dual geometry at asymptotically large $\tau$ turns out to be the scaling solution \cite{Janik:2005zt}
\beq
ds^2=\frac{1}{z^2}\left(-e^{a(u)}d\tau^2+\tau^2 e^{b(u)}d\zeta^2 + e^{c(u)} d\vec{x}_T^2+dz^2\right)\,. \label{scale}
\eeq
The scaling variable is defined by $u\equiv z/\tau^{s/4}$ where $s$ is a parameter required to satisfy $0\le s\le 4$ from the positivity constraint of the energy density.
 The Bjorken solution $\varepsilon\sim 1/\tau^{4/3}$ corresponds to $s=4/3$ and
\beq
a(u)\approx \frac{(1-\frac{e_0}{3}u^4)^2}{(1+\frac{e_0}{3}u^4)}\,, \quad b(u)\approx c(u)\approx 1+\frac{e_0}{3}u^4\,, \label{ja}
\eeq
  where $e_0$ is a dimensionful normalization constant. (\ref{ja}) satisfies the Einstein equation up to terms subleading (at fixed $u$) in inverse powers of $\tau$ which are related to the viscous effects \cite{Nakamura:2006ih,Janik:2006ft}.
It was shown in \cite{Janik:2005zt} that within the range $0<s<4$, the value $s=4/3$ is the only acceptable choice after requiring that there are no singularities in the bulk. The boundary value $s=4$ was not studied in \cite{Janik:2005zt} because, as the authors pointed out, their method to solve the Einstein equation (expansion in negative powers of $\tau$ at fixed $u$) fails when $s=4$. What happens at $s=4$ is that $u=z/\tau$ is dimensionless, and therefore at fixed $u$ negative powers of $\tau$ cannot appear. In other words, one has to solve the Einstein equation exactly. Furthermore, it is interesting to note that not only $e_0$ becomes dimensionless when $s=4$, but also it becomes a fixed number due to nonlinearity of the Einstein equation as shown below.

The reason why we are interested in the special value $s=4$ is that, if a solution with $s=4$ indeed exists, then it is dual to a hydrodynamic flow $\varepsilon \sim \frac{1}{\tau^4}$ which has the same $\tau$-dependence as (\ref{rt}). Unfortunately, we have not been able to solve the Einstein equation for $s=4$ exactly in a closed form. Instead, we performed a perturbative expansion of $a(u)$, etc. in Eq.~\eqref{scale} to high orders. The result is
\begin{eqnarray}
a(u)&=&-u^4-\frac{2}{3}u^6-u^8-\frac{4}{3} u^{10} -2 u^{12} -\frac{65}{21}u^{14} -\frac{179}{36}u^{16}-\frac{172}{21}u^{18}-\frac{8681}{630}u^{20}+\cdots\,, \label{bow} \\
b(u)&=&3u^4+\frac{10}{3}u^6+3u^8+4 u^{10} +\frac{22}{3} u^{12} +\frac{95}{7}u^{14}+\frac{287}{12}u^{16}+\frac{880}{21}u^{18}+\frac{15761}{210}u^{20}
+\cdots\,, \nonumber\\
c(u)&=&-u^4-\frac{4}{3}u^6-2 u^8-\frac{10}{3} u^{10} -\frac{17}{3} u^{12} -\frac{208}{21}u^{14} -\frac{635}{36}u^{16}-\frac{223}{7}u^{18}-\frac{36661}{630}u^{20}+\cdots\,. \nonumber
\end{eqnarray}
In fact, the expansion can be carried out up to all orders \cite{de Haro:2000xn}. Remarkably, to lowest order the coefficient of $u^4$ in $a(u)$ is arbitrary, but once we go to higher orders the nonlinearity of the Einstein equation selects the coefficient to be $-1$. Assuming that (\ref{bow}) represents the near-boundary behavior of a well-defined solution, we read off the energy density in field theory
\beq
\varepsilon=\frac{N_c^2}{2\pi^2}\frac{1}{\tau^4}\,.  \label{nc}
\eeq
 Once again, we find that the coefficient of $1/\tau^4$ is uniquely determined, this time due to the nonlinearity of the Einstein equation. Moreover, the energy momentum tensor has the form
 \beq
 \varepsilon=T^{\tau\tau}=-T^{xx}=-T^{yy}=\frac{\tau^2}{3}T^{\zeta\zeta}\,,
 \eeq
  which is exactly the same as that for the unorthodox Bjorken flow  \cite{Hatta:2014gga}.
To make the comparison clearer, we note that by using Eq.~\eqref{beta}
\beq
\frac{N_c^2}{2\pi^2} =C = (\beta\hat{\tau}_\pi)^4 \,.% =\bar r_\pi^4 \tau_\pi^4 \varepsilon\,.
\eeq
 We then use the known results at strong coupling \cite{Baier:2007ix,Bhattacharyya:2008jc}
  \beq
  \varepsilon=\frac{3\pi^2N_c^2T^4}{8}\,, \qquad \tau_\pi=\frac{2-\ln 2}{2\pi T}\,,
  \eeq
 to rewrite  Eq.~\eqref{nc} as
 \beq
 \varepsilon=\frac{(\beta\hat{\tau}_\pi)^4}{\tau^4}\,,  \qquad \textrm{with} \quad \beta=\frac{2(4/3)^{1/4}}{2-\ln 2} \approx 1.64\,,
 \eeq
%\beq
%\tau_\pi=\frac{1}{a\pi T} \left(\frac{4}{3}\right)^{1/4}
%\eeq
 which is rather close to Eq.~\eqref{rt}, though of course the agreement is not expected since one is comparing strongly coupled and weakly coupled theories.

Unfortunately, we have not been able to resum the series~\eqref{bow} in a closed form and this prohibits us from studying the possible singularities of the geometry in the bulk. Moreover, the radius of convergence for the series \eqref{bow} seems to be decreasing towards zero. Therefore, at the moment we cannot draw a conclusion about the existence of solutions of the form (\ref{nc}), although it would be very interesting to have one in view of the results in kinetic theory (\ref{rt}) and hydrodynamics (\ref{ortho}). We leave this problem to future work.

%%%%%%%%%%%%%%%%%%%%%%%%%%%%%%%%%%%%%%%%%%%%%%%%%%%%%%%%%%%%%%%%%%%%%%%%%%%%%%%%%%%%%%%%%%%
\section{Conclusions}
\label{sec:concl}
%%%%%%%%%%%%%%%%%%%%%%%%%%%%%%%%%%%%%%%%%%%%%%%%%%%%%%%%%%%%%%%%%%%%%%%%%%%%%%%%%%%%%%%%%%%

In this paper, we have derived analytic solutions of the Boltzmann equation in the RTA for conformally invariant systems. In the `Hubble' flow case, we have shown that the solutions of the kinetic (Boltzmann) and hydrodynamic (Israel-Stewart) equations are essentially the same, in that they give the identical conserving energy momentum tensor. Most of the previous studies have found systematic differences between the  solutions of the kinetic and hydrodynamic equations, the former is considered to be more fundamental. However, in the present case they agree exactly.

We then considered the Boltzmann equation in the boost-invariant setup relevant to the final state of heavy-ion collisions. At asymptotically large times, the equation selects the Bjorken solution $\varepsilon \sim 1/\tau^{4/3}$. However, we have pointed out that in the conformally symmetric case a novel solution of the form $\varepsilon = C/\tau^4$ exists, and we precisely determined the normalization $C$ which is not an arbitrary parameter. Very interestingly a solution with the same $\tau$-dependence was previously found as an exact solution of second order hydrodynamics. The existence of this solution is also suggested by fluid-gravity correspondence. But here our argument is not adequate, and more work is needed to firmly establish (or exclude) such a solution of the Einstein equation.

We conclude by mentioning some possibilities to explore from our findings which might have potential applications in more realistic scenarios such as in high energy nuclear collisions.  At early times after the collision between highly energetic nuclei, the produced plasma of quarks and gluons reaches high temperatures with almost vanishing chemical potential such that it is approximately conformal. This tiny plasma is far away from equilibrium due to the violent acceleration in all the spatial directions and thus, the 3D Hubble solution discussed in Sect.~\ref{hubblesec} offers the possibility to model at this stage the system. In addition, the 3D Hubble solution can encode the early-time information about the flow configuration and energy density behavior $\sim\tau_r^{-4}$. This scaling of the energy density at the beginning of the expansion might be connected with the unorthodox Bjorken flow solution whose energy density decays faster $\sim\tau^{-4}$ than the standard Bjorken solution~$\sim{\tau^{-4/3}}$~\cite{Bjorken:1982qr}. Therefore, within this scenario the lifetime of the fireball would be shorter since the unorthodox Bjorken flow predicts that the system reaches faster the freeze-out temperature. This possibility can be studied  nowadays by performing a systematic fine-tuning analysis of the freeze-out temperature, the initial time when the numerical hydrodynamical simulation starts to run and other parameters such as the transport coefficients.

\section*{Acknowledgements}
We thank Gabriel Denicol and Jorge Noronha for collaboration in the early stage of this work. Y.~H. thanks Shinji Mukohyama for helpful conversations.
M.~M. would like to thank the Physics Department of the Universidade de S\~ao Paulo for hosting him during the early stages of this project.
 B.~X. wishes to thank Feng Yuan and the nuclear theory group at the Lawrence Berkeley National Laboratory for hospitality and support during his visit when this work is finalized.
M.~M. is supported by the U.~S. Department of Energy, Office of Science, Office of Nuclear Physics under Award No. DE-SC0004286, and also by a bilateral scientific exchange program between the Office of Sponsored Research at The Ohio State University and FAPESP.

\appendix

\section{Another derivation of (\ref{aa})}
\label{app}

In order to be more confident with the unfamiliar number $\beta=1.27672$ found in (\ref{aa}), let us derive it from a slightly different perspective.
Doing integration by parts in the right-hand side of (\ref{care}) and keeping contributions only from $\tau'=\tau$, we get
\beq
 \text{Eq.}(\ref{care})&\sim& \frac{C}{\tau^{4-\beta}} -\frac{\beta C}{2}\int^\tau_{\tau_0} \frac{d\tau'}{\tau'^{5-\beta}} \left(\frac{\tau'^2}{\tau^2} + \frac{\arctan \sqrt{\frac{\tau^2}{\tau'^2}-1}}{\sqrt{\frac{\tau^2}{\tau'^2}-1}}\right) \nn
&\sim& \frac{C}{\tau^{4-\beta}}\left(1+\frac{\beta}{2(2-\beta)} -\frac{\beta}{2}\int^{\sqrt{\tau^2/\tau_0^2-1}}_0 dx (x^2+1)^{1-\beta/2} \arctan x \right)\,,
\eeq
 where we changed variables as $x=\sqrt{\tau^2/\tau'^2 -1}$. The $x$-integral is convergent at $x\to \infty$ when $\beta>3$. We thus temporarily assume $\beta>3$ and send the upper limit to infinity. Then the integral can be done exactly
 \beq
 \frac{C}{\tau^{4-\beta}}\left\{1+\frac{\beta}{2(2-\beta)} + \frac{\beta\sqrt{\pi}}{8}\Gamma\left(2-\frac{\beta}{2}\right)
 \left(\frac{\Gamma\left(\frac{\beta-3}{2}\right)}{\sin \frac{\beta\pi}{2}}+ \frac{\ _3F_2 \left(\frac{1}{2},1,1;\frac{3}{2},3-\frac{\beta}{2};1\right)}{\Gamma(\frac{3}{2})\Gamma(3-\frac{\beta}{2})}
 \right) \right\}\,. \label{zero}
 \eeq
We then analytically continue the result to $\beta<3$ and look for the zero of (\ref{zero}). We find $\beta\approx 1.27672$ in perfect agreement with the previous method.

%%%%%%%%%%%%%%%%%%%%%%%%%%%

%%%%%%%%%%%%%%%%%%%%%%%%%%%%

\end{document}